\documentclass[conference]{IEEEtran}
\IEEEoverridecommandlockouts

\usepackage{cite}
\usepackage{amsmath,amssymb,amsfonts}
\usepackage{booktabs}
\usepackage{microtype}
\usepackage{enumitem}
\usepackage[vlined,ruled,linesnumbered]{algorithm2e}
\usepackage{threeparttable}
\usepackage{multirow}
\usepackage{graphicx}
\usepackage{placeins}
\usepackage{url}
\usepackage{tabularx}
\usepackage{array}
\usepackage{makecell}
\usepackage{textcomp}

\newcommand{\StrideEmbed}{\text{StrideEmbed}}
\newcommand{\BudgetResolver}{\text{BudgetResolver}}
\newcommand{\TopK}{\text{TopK}}
\newcommand{\Expand}{\text{Expand}}
\newcommand{\Gather}{\text{Gather}}
\newcommand{\WPool}{\text{WPool}}
\newcommand{\Enc}{\text{Enc}}
\newcommand{\Pool}{\text{Pool}}
\newcommand{\MLP}{\text{MLP}}

\newcommand{\methodname}{NetCoFi}

\newcommand{\Lb}{\text{L}_\text{b}}
\newcommand{\ds}{\text{d}_\text{s}}

\newcommand{\keff}{\text{k}_{\text{eff}}}
\newcommand{\reff}{\text{r}_{\text{eff}}}
\newcommand{\reffused}{\text{r}_{\text{eff}}^{\text{used}}}

\newcommand{\sel}{\text{sel}}

\SetAlCapHSkip{0.5em}
\SetInd{0.2em}{1em}
\SetKwInput{Input}{Input}
\SetKwInput{Output}{Output}
\SetKwComment{Comment}{/* }{ */}
\SetNlSty{textnormal}{}{:}

\begin{document}

\title{NetCoFi: A Budget-Aware Coarse-Fine Dual-Branch Model for Encrypted Traffic Classification}

\author{
\IEEEauthorblockN{
Zhengyi Ma\textsuperscript{1,\textdagger},
Qiming Yu\textsuperscript{2,\textdagger},
Chungang Lin\textsuperscript{3,4},
Wentao Cui\textsuperscript{1},
Zijia Song\textsuperscript{2},
Guozheng Yang\textsuperscript{2,*}
}

\IEEEauthorblockA{
\textsuperscript{1}College of Computer Science and Technology,
National University of Defense Technology, Changsha 410073, China\\
\textsuperscript{2}College of Electronic Engineering,
National University of Defense Technology, Hefei 230007, China\\
\textsuperscript{3}Institute of Computing Technology,
Chinese Academy of Sciences, China\\
\textsuperscript{4}University of Chinese Academy of Sciences, China
}
}

\maketitle

\begingroup
\renewcommand{\thefootnote}{\fnsymbol{footnote}}
\footnotetext[2]{These authors contributed equally to this work.}
\footnotetext[1]{The corresponding author is Guozheng Yang.}
\endgroup

\begin{abstract}
Encrypted traffic classification is critical for network management and security, but practical deployment must balance classification accuracy and inference cost. Existing models typically process fixed flow windows with limited control over computation after training, which can limit the accuracy–efficiency trade-off under constrained resources. To address this problem, we present NetCoFi, a budget-aware classifier that allocates computation at coarse and fine granularities. Using byte representations and observable packet metadata, NetCoFi selects token neighborhoods for fine-grained encoding while compressing the full window into complementary coarse context. Multi-budget training enables the same model to operate at different routing budgets without retraining. Evaluation on six public datasets shows that NetCoFi achieves state-of-the-art classification performance when trained from scratch. Under low-budget settings, NetCoFi improves average Macro-F1 by 7.86 percentage points across three datasets, compared with a same-depth static baseline. In a DPDK–GPU traffic processing pipeline, it increases mean batch throughput from 26.42 to 43.42 Mb/s, a 1.64× improvement over NetMamba.
\end{abstract}


\section{Introduction}
Efficient encrypted traffic classification is an important capability for network management and security monitoring, supporting application identification and traffic analysis~\cite{cite_bg2,ml_survey2022,cite_bg1}. As TLS and encrypted tunnels become widely used, plaintext payload semantics are increasingly unavailable, forcing classifiers to infer traffic classes from observable byte sequences and packet-level metadata~\cite{cite_bg1,sok_enigma}. At the same time, practical deployment must meet latency and throughput constraints. 
This raises a key question: how to ensure that the classifier, given limited inference resources, can retain sufficient traffic information to make accurate predictions while avoiding computational waste?

Recent studies have improved encrypted traffic classification through advances in representation learning and sequence modeling.
End-to-end models learn features directly from traffic data and reduce reliance on manual feature design~\cite{fsnet}.
Pretrained models use unlabeled traffic to learn representations that can be adapted to downstream classification tasks~\cite{etbert,yatc,flowmae}.
More recently, Mamba-based classifiers have improved backbone efficiency with state-space encoders~\cite{netmamba,pat}.
These encoders reduce the computational complexity of long-sequence modeling compared with standard self-attention~\cite{mamba,dao2024ssd}.

However, many classifiers still process fixed traffic windows along fixed computation paths.
Once trained, these models use a fixed computation path for every prediction, regardless of input complexity or available resources.
Shortening the input window can reduce this cost, but it also removes traffic information before the model can assess its value.
A more flexible alternative is to adjust the amount of computation within the window instead of relying only on a shorter input.

Such an approach must balance local detail with global context.
Regions within a flow window may contribute unequally to classification.
Processing all regions at the same granularity can waste computation on those that provide little useful information.
However, a region that is less informative on its own may still provide context for other parts of the window.
Discarding it entirely can therefore remove information that supports the final prediction.
NetCoFi therefore concentrates fine-grained processing on informative regions, while retaining a global summary of the full window.

\begin{figure}[!t]
    \centering
    \includegraphics[width=0.90\linewidth]{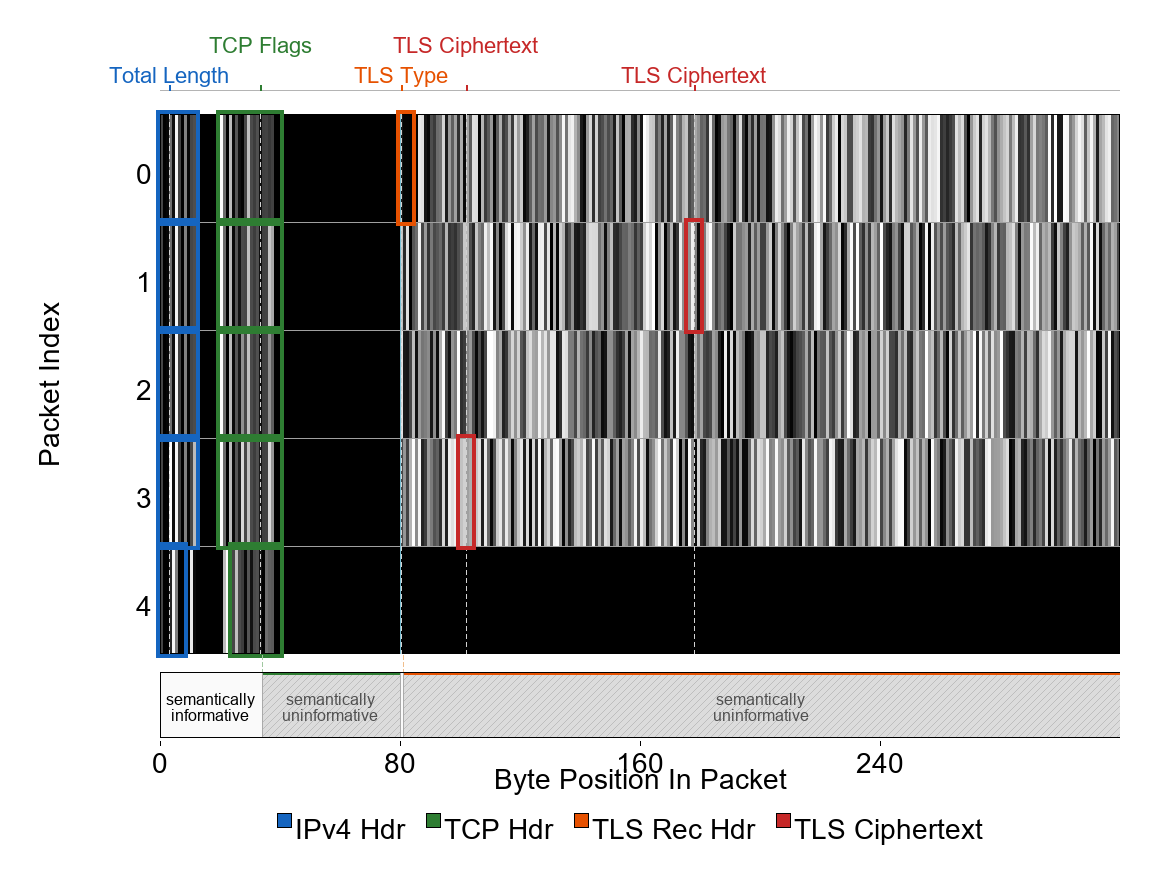}
    \caption{Illustration of sparse discriminative regions in encrypted traffic and the localized byte-level semantics that motivate compute-aware routing.}
    \label{fig:motivation}
\end{figure}

We propose NetCoFi, a budget-aware coarse-fine dual-branch classifier for fixed-window encrypted traffic classification.
It estimates token importance from byte representations and token-aligned side features derived from packet metadata.
The fine branch encodes selected token neighborhoods in detail.
Including neighboring tokens preserves local context around the selected positions.
The coarse branch compresses the full window, including regions not selected by the fine branch.
It provides broader context without applying fine-grained encoding to every region.
The two branch representations are then fused for classification.

Changing the routing budget also changes the information available to the classifier.
A model trained only at full budget may lose accuracy when less detail is retained.
We therefore train NetCoFi under multiple routing budgets with shared parameters.
This exposes the model to different levels of detail during training.
At inference time, the user can choose a budget according to the available computational resources.
The same model then provides different accuracy--efficiency choices without retraining.

We evaluate NetCoFi on six public benchmarks and in a DPDK--GPU traffic processing pipeline.
In the low-budget comparison, it outperforms a same-depth static token-reduction baseline.
The average Macro-F1 gain is 7.86 percentage points across three datasets.
In the deployment pipeline, its mean batch throughput reaches 43.42~Mb/s, compared with 26.42~Mb/s for NetMamba.
This throughput comparison includes flow aggregation and feature construction, rather than measuring model inference alone.

In summary, this paper makes the following contributions:
\begin{enumerate}
    \item We propose NetCoFi, a coarse-fine dual-branch model for encrypted traffic classification. It allocates fine-grained computation to informative local regions, while preserving global context through a coarse branch.
    \item We introduce a budget-aware training strategy. It allows one shared model to operate under different routing budgets without retraining. This provides a flexible accuracy–efficiency trade-off under different resource constraints.
    \item We evaluate NetCoFi on six public datasets and in a DPDK–GPU deployment pipeline. NetCoFi achieves competitive classification performance, stronger low-budget accuracy, and higher deployment throughput than NetMamba.
\end{enumerate}

\section{Related Work}
\subsection{Encrypted Traffic Classification}

Encrypted traffic classification typically relies on side channel information, such as packet length, direction, header fields, and handshake metadata, rather than plaintext payloads~\cite{cite_bg2,ml_survey2022,sok_enigma,cite_bg1}. Early methods such as AppScanner and FlowPrint use statistical or side-channel fingerprints for application identification~\cite{appscanner,flowprint}. Deep learning methods further reduce the dependence on handcrafted features by learning discriminative representations directly from traffic data. Deep Packet, FS-Net, and TFE-GNN model raw bytes or structural relationships in traffic, allowing classifiers to capture both local patterns and longer-range dependencies~\cite{deeppacket,fsnet,tfegnn}.

Recent studies increasingly use self-supervised learning and pretraining. PERT and ET-BERT learn traffic representations from large amounts of unlabeled data~\cite{pert,etbert}. YaTC and Flow-MAE further introduce masked modeling and autoencoding objectives~\cite{yatc,flowmae}. These methods improve traffic representation quality, but their main goal is classification performance or transferability. They do not explicitly control inference cost after training.

\subsection{Efficient Encrypted Traffic Classification}

Existing studies reduce the cost of encrypted traffic classification through lightweight architectures and sequence modeling. Compact temporal models and lightweight graph encoders reduce model complexity~\cite{clstm_efficient,lightweight_graph}. However, these methods generally retain a fixed computation path during inference.

State-space models provide another direction for efficient sequence modeling. Mamba reduces the complexity of long-sequence modeling compared with standard self-attention~\cite{mamba,dao2024ssd}. NetMamba and PaT apply Mamba-style architectures to encrypted traffic analysis~\cite{netmamba,pat}. Their efficiency mainly comes from the backbone architecture. The input window is still processed with a largely fixed computational workload. In contrast, NetCoFi focuses on how computation is allocated within a fixed traffic window under different routing budgets.

\subsection{Adaptive Computation and Budget-Aware Routing}

Conditional computation dynamically allocates resources according to input content. Existing vision and language models use token pruning or adaptive routing to reduce redundant computation~\cite{ryoo2021tokenlearner,rao2021dynamicvit,liang2022evit,yin2022avit,bolya2023tome,mixture_of_depths,dtrnet}. Other methods dynamically select layers or model paths~\cite{teerapittayanon2016branchynet,xin2020deebert,graves2016act,wang2018skipnet,fedus2021switch}. These approaches avoid applying the same amount of computation to all inputs or token regions. As a result, they can reduce inference cost while preserving important information, and provide a more flexible accuracy--efficiency trade-off under different resource constraints.

Dynamic computation has received less attention in encrypted traffic classification. FastFlow dynamically determines how many packets should be observed, while pForest switches models according to different flow stages~\cite{fastflow,pforest}. Their decisions are mainly made at the packet, flow-stage, or model-path level. By contrast, NetCoFi allocates computation among token regions within a fixed flow window. It combines fine-grained processing of informative regions with coarse-grained global context under a controllable routing budget.
\section{Method}
\label{sec:method}

\begin{figure*}[htbp]
  \centering
  \includegraphics[width=0.98\textwidth]{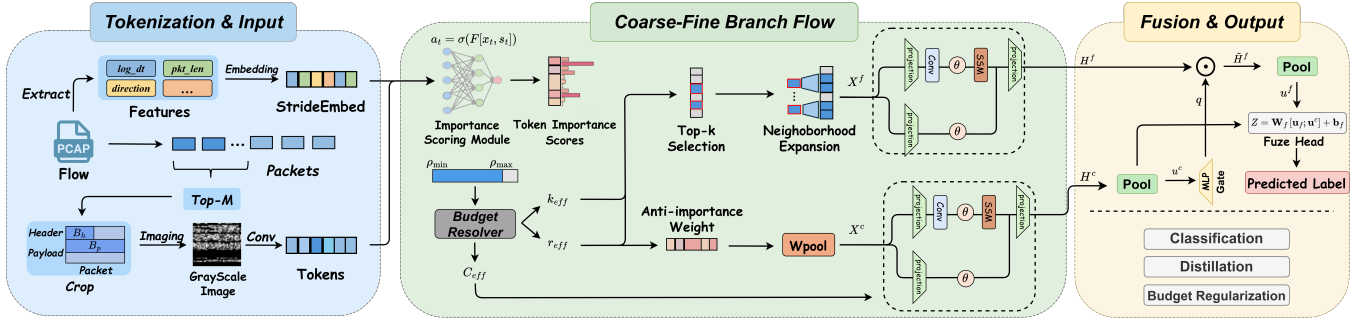}
  \caption{\methodname{} architecture. Budget-aware routing constructs fine-grained token neighborhoods and compressed coarse context. Two Mamba branches encode these representations, which are combined through cross-gated fusion.}
  \label{fig:architecture}
\end{figure*}

\subsection{Method Overview}
\label{sec:method_overview}

NetCoFi is a budget-aware coarse-fine dual-branch classifier designed to concentrate computation on informative tokens while preserving complementary global context. As illustrated in Fig.~\ref{fig:architecture}, the input byte stream is first tokenized into stride tokens $X_0\in\mathbb{R}^{N\times d}$. From $X_0$ and token-aligned side features $s$, NetCoFi computes token importance scores $\mathbf{a}=[a_1,\dots,a_N]$. The user-specified routing budget $\rho$ is then resolved into $(k_{\mathrm{eff}}, r_{\mathrm{eff}}, C_{\mathrm{eff}})$, which determine the seed-token budget, the budget-resolved neighborhood radius, and the coarse-token budget. The fine branch processes the selected seed tokens together with their budget-resolved neighborhoods, whereas the coarse branch compresses the full sequence into complementary coarse context through anti-importance weighted pooling. The two branch representations are finally combined by cross-gated fusion to produce logits $z\in\mathbb{R}^C$.

Algorithm~\ref{alg:NETCOFI} summarizes the NetCoFi inference forward pass used throughout our experiments.

\begin{algorithm}[t]
\caption{NetCoFi Inference Forward Pass}
\label{alg:NETCOFI}
\small

\Input{%
  flow-window byte array $\mathbf{b} \in \{0,\ldots,255\}^{\Lb}$,
  token-aligned side features $s \in \mathbb{R}^{N\times \ds}$,
  routing budget $\rho \in [\rho_{\min}, \rho_{\max}]$
}

\Output{%
  logits $z \in \mathbb{R}^{C}$, predicted label $\hat{y}$
}
\BlankLine

$X_0 \leftarrow \StrideEmbed(\mathbf{b})$\;
$\mathbf{a} \leftarrow [\sigma(f([\mathrm{Norm}(x_t), \mathrm{Norm}(s_t)]))]_{t=1}^{N}$\;
$(\keff, \reff, C_{\mathrm{eff}}) \leftarrow \BudgetResolver(\rho)$\;
$I_{\mathrm{seed}} \leftarrow \TopK(\mathbf{a}, \keff)$\;

\If{\textnormal{EnableExpand}}{
  $\reffused \leftarrow \reff$\;
}
\Else{
  $\reffused \leftarrow 0$\;
}

$I_{\sel} \leftarrow \Expand(I_{\mathrm{seed}}, \reffused)$\;
$X^f \leftarrow \Gather(X_0, I_{\sel})$\;
$X^c \leftarrow \WPool(X_0, 1-\mathbf{a}, C_{\mathrm{eff}})$\;

$H^f \leftarrow \Enc^f(X^f)$\;
$H^c \leftarrow \Enc^c(X^c)$\;
$u^c \leftarrow \Pool(H^c)$\;

\If{\textnormal{EnableCrossGate}}{
  $q \leftarrow \sigma(\MLP_{\mathrm{gate}}(u^c))$\;
  $\tilde{H}^f \leftarrow H^f \odot q$\;
}
\Else{
  $\tilde{H}^f \leftarrow H^f$\;
}

$u^f \leftarrow \Pool(\tilde{H}^f)$\;
$z \leftarrow W_f [u^f; u^c] + b_f$\;
$\hat{y} \leftarrow \arg\max_c z_c$\;

\textbf{return} $z, \hat{y}$\;

\end{algorithm}

\subsection{Input Representation and Side Features}
\label{sec:method_tokenization}
\label{sec:method_input}

We consider supervised encrypted traffic classification on a labeled dataset $\mathcal{D}$. Each sample consists of an input $x=(\mathbf{b},s)$ and a label $y\in\{1,\dots,C\}$. Here, $\mathbf{b}$ is a fixed-length byte sequence, and $s$ contains token-aligned side features.

For each flow, we extract the first $M$ packets. Header and payload bytes are truncated or padded to $B_h$ and $B_p$ bytes per packet. We concatenate these regions in packet order to form $\mathbf{b}\in\{0,\dots,255\}^{L_b}$, where $L_b=M(B_h+B_p)$. The default configuration uses $M=5$, $B_h=80$, and $B_p=240$, giving $L_b=1600$. 

A one-dimensional convolution with stride $p$ projects the byte sequence into token embeddings:
\begin{equation}
X_0=\mathrm{StrideEmbed}(\mathbf{b})\in\mathbb{R}^{N\times d},
\label{eq:stride_embed}
\end{equation}
where $N=L_b/p$ and $d$ is the embedding dimension.

Each token $x_t$ is paired with a side feature vector $s_t\in\mathbb{R}^{d_s}$, yielding $s=[s_1,\dots,s_N]$. Packet and region descriptors are derived from the byte positions covered by each token. These descriptors identify the corresponding packet, header or payload region, and relative position. Table~\ref{tab:side_token_feats} summarizes the side features. Continuous features are normalized using training-set statistics. The same transformation is applied to validation and test data.

\begin{table}[htbp]
\centering
\caption{Token-aligned side features used in NetCoFi.}
\label{tab:side_token_feats}
\renewcommand{\arraystretch}{1.08}
\resizebox{\linewidth}{!}{%
\begin{tabular}{c|l|l}
\hline
\textbf{\#} & \textbf{Name} & \textbf{Description} \\
\hline
1  & log\_dt            & $\log(1+\Delta t)$ inter-packet gap \\
2  & pkt\_len\_norm     & normalized packet length \\
3  & direction\_bin     & direction indicator (uplink/downlink) \\
4--11 & syn, ack, rst, fin, psh, urg, ece, cwr & TCP flag indicators \\
12 & is\_tcp            & TCP indicator \\
13 & delta\_len\_norm   & normalized length difference to previous packet \\
14 & dir\_switch        & whether direction switches vs previous packet \\
15 & pkt\_pos\_norm     & normalized packet index within the prefix \\
16 & tok\_pos\_norm     & normalized token index within a packet \\
17 & is\_header\_token  & whether the token belongs to header region \\
18 & header\_pos\_norm  & normalized position within header region \\
19 & log\_local\_rate   & $\log(1+\mathrm{len}/\Delta t)$ local rate proxy \\
20 & dt\_z\_dataset     & dataset-level z-score of log\_dt \\
\hline
\end{tabular}%
}
\end{table}

\subsection{Budgeted Coarse-Fine Routing}
\label{sec:method_core}
\label{sec:method_c3}

Rather than allocating computation uniformly across all stride tokens, NetCoFi learns an importance map and uses it to construct complementary fine and coarse sequences.

\paragraph{Importance scoring}
Given $X_0=[x_1,\dots,x_N]$ and token-aligned side features $s=[s_1,\dots,s_N]$, the importance scores are computed as
\begin{equation}
a_t = \sigma\big(f([\mathrm{Norm}(x_t),\mathrm{Norm}(s_t)])\big)\in(0,1),\quad t=1,\dots,N,
\label{eq:importance}
\end{equation}
where $f(\cdot)$ is a lightweight multi-layer perceptron applied to the concatenation of normalized projections of the token embedding and the side feature. These scores are shared by both branches: they drive fixed-$k$ seed selection in the fine branch and anti-importance weighting in the coarse branch.

\paragraph{Budget resolver}
We use a routing budget
$\rho\in[\rho_{\min},\rho_{\max}]\subseteq[0,1]$.
The default interval is $[0,1]$.
The user specifies $\rho$ at inference time to control the routing workload.
This value is not a literal fraction of raw tokens retained.
The full routing configuration corresponds to $\rho=1.0$.

A deterministic resolver maps $\rho$ to
$(k_{\mathrm{eff}},r_{\mathrm{eff}},C_{\mathrm{eff}})$.
These variables determine the seed-token count, neighborhood radius,
and coarse-token count, respectively.
Let $K_0$ and $C_0$ denote the maximum seed-token and coarse-token counts
at $\rho=1.0$.
The main configuration uses $K_0=64$ and $C_0=100$.
The token counts follow
\begin{equation}
k_{\mathrm{eff}}=\mathrm{round}(K_0\cdot\rho),\qquad
C_{\mathrm{eff}}=\max\{1,\mathrm{round}(C_0\cdot\rho)\}.
\label{eq:budget_map}
\end{equation}
We set $r_{\mathrm{cap}}=\min\{\textit{focal\_r},\textit{budget\_r\_max}\}$.
The neighborhood radius follows
\begin{equation}
r_{\mathrm{eff}}=
\begin{cases}
r_{\mathrm{cap}}, & \rho \ge 0.75,\\
\max\{1,r_{\mathrm{cap}}-1\}, & 0.50 \le \rho < 0.75,\\
0, & \rho < 0.50.
\end{cases}
\label{eq:reff_schedule}
\end{equation}

The main configuration uses $\textit{focal\_r}=\textit{budget\_r\_max}=2$.
With neighborhood expansion enabled, $r_{\mathrm{eff}}^{\mathrm{used}}=r_{\mathrm{eff}}$.
Otherwise, $r_{\mathrm{eff}}^{\mathrm{used}}=0$.

\paragraph{Fine-token construction}
Using this routing tuple, the fine branch first selects exactly $k_{\mathrm{eff}}$ seed tokens via fixed-$k$ routing:
\begin{equation}
I_{k_{\mathrm{eff}}} = \mathrm{TopK}(\mathbf{a}, k_{\mathrm{eff}}).
\label{eq:topk}
\end{equation}
Each selected seed is then expanded in token-index space using the applied neighborhood radius $r_{\mathrm{eff}}^{\mathrm{used}}$, i.e., over adjacent stride tokens rather than raw bytes. This yields the fine-token index set
\begin{equation}
I^f = \bigcup_{i\in I_{k_{\mathrm{eff}}}} \{\,j \mid |j-i|\le r_{\mathrm{eff}}^{\mathrm{used}},\; 1\le j\le N\,\}.
\label{eq:fine_expand}
\end{equation}
The fine-token sequence is then formed by gathering the expanded neighborhoods from $X_0$ while preserving the implementation order used by the fine branch:
\begin{equation}
X^f = \mathrm{Gather}(X_0, I^f)\in\mathbb{R}^{N_f\times d},
\label{eq:fine_gather}
\end{equation}
Here $N_f$ is the gathered input length processed by the fine encoder.
It includes neighborhood expansion and is not a deduplicated count of unique tokens.

\paragraph{Coarse-token construction}
To preserve complementary global context while concentrating fine-grained computation on the routed region, we pool the full token stream into $N_c=C_{\mathrm{eff}}$ coarse tokens. Let $\mathbf{w}=1-\mathbf{a}$ denote the anti-importance weights. A weighted pooling operator $\mathrm{WPool}(\cdot)$ aggregates information along the sequence dimension and produces a fixed-length token sequence:
\begin{equation}
X^c = \mathrm{WPool}(X_0,\mathbf{w},N_c)\in\mathbb{R}^{N_c\times d}.
\label{eq:coarse_pool}
\end{equation}
$\mathrm{WPool}$ is implemented as weighted adaptive average pooling over contiguous bins, where each output token is the weighted mean of the tokens assigned to that bin.

\subsection{Dual-Branch Encoding and Cross-Gated Fusion}
\label{sec:method_fusion}
\label{sec:method_c2}

After routing constructs $X^f$ and $X^c$, NetCoFi encodes the two sequences with separate Mamba-style encoders:
\begin{equation}
H^f = \mathrm{Enc}^f(X^f)\in\mathbb{R}^{N_f\times d},\quad
H^c = \mathrm{Enc}^c(X^c)\in\mathbb{R}^{N_c\times d}.
\label{eq:dual_enc}
\end{equation}

The coarse summary generates a channel-wise gate for the fine states. The pooled branch outputs are then concatenated for classification:
\begin{equation}
\begin{aligned}
u^c &= \mathrm{Pool}(H^c),\\
q &= \sigma(\mathrm{MLP}_{\mathrm{gate}}(u^c))\in(0,1)^d,\\
\tilde{H}^f &= H^f \odot q,\\
u^f &= \mathrm{Pool}(\tilde{H}^f),\\
z &= W_f [u^f;u^c] + b_f.
\end{aligned}
\label{eq:cross_gate}
\end{equation}
This cross-gated fusion preserves the fine branch as the discriminative local pathway while allowing coarse global context to condition the final fine-token summary before classification.

\subsection{Multi-Budget Training Objective}
\label{sec:method_budget}

We train one shared model $f_\theta$ across multiple routing budgets.
For each mini-batch $\mathcal{B}_t$, we sample the two endpoint budgets and $Q$ intermediate budgets:
\begin{equation}
\begin{aligned}
\mathcal{S}_t
&=\{\rho_{\max}, \rho_{\min}, u_{t,1}, \dots, u_{t,Q}\}, \\
u_{t,q}
&\sim \mathrm{Uniform}[\rho_{\min}, \rho_{\max}],
\end{aligned}
\label{eq:sandwich_set_theory}
\end{equation}

To further stabilize low-budget predictions, we optionally use the full-budget model as an online teacher. Let $f_\theta(\cdot;1.0)$ denote the teacher prediction. For each reduced budget $\rho<1$, we apply
\begin{equation}
\begin{aligned}
\mathcal{L}_{\mathrm{kd}}^{(\rho)}
=
T_{\mathrm{kd}}^2 \, \mathrm{KL}\Big(
&\mathrm{softmax}(f_\theta(x;1.0)/T_{\mathrm{kd}}) \\
&\;\|\;
\mathrm{softmax}(f_\theta(x;\rho)/T_{\mathrm{kd}})
\Big),
\end{aligned}
\label{eq:kd_objective}
\end{equation}
which encourages lower-budget predictors to match the predictive distribution of the full-budget model. We stop the gradient through the full-budget teacher distribution. We set $\mathcal{L}_{\mathrm{kd}}^{(1.0)}=0$ for the full-budget pass. In addition, because routing is controlled by the explicit routing budget, we regularize the importance scores so that the realized routing tendency remains calibrated to the requested budget. Let $\mathbf{a}(x)=[a_1,\dots,a_N]$ be the importance scores from Eq.~\eqref{eq:importance}, and let $N'$ denote the effective token length to which the routing budget is applied ($N'=N$ in the default setting without pre-subsampling). We define
\begin{equation}
\rho^{\star}(\rho)=\frac{k_{\mathrm{eff}}}{N'},
\qquad
\mathcal{L}_{\mathrm{budget}}^{(\rho)}
=
\left(
\frac{1}{N'}\sum_{t=1}^{N'} a_t
- \rho^{\star}(\rho)
\right)^2.
\label{eq:budget_regularizer}
\end{equation}
This budget-regularization term encourages the mean importance level to track the target seed-token ratio implied by $\rho$, thereby reducing routing drift across operating points.

The final objective is
\begin{equation}
\begin{aligned}
\min_{\theta} \
\mathbb{E}_{(x,y)} \mathbb{E}_{\rho}
\Big[\,
&\ell(f_\theta(x;\rho), y)
+ \lambda_{\mathrm{kd}} \mathcal{L}_{\mathrm{kd}}^{(\rho)} \\
&+ \lambda_{\mathrm{budget}} \mathcal{L}_{\mathrm{budget}}^{(\rho)}
\Big].
\end{aligned}
\label{eq:final_objective_theory}
\end{equation}
The expectation over $\rho$ is approximated by averaging over the sandwich budget set $\mathcal{S}_t$ in Eq.~\eqref{eq:sandwich_set_theory}.

Training exposes the shared model to both endpoint and intermediate budgets. At inference time, the routing budget can be changed without retraining. Section IV-C evaluates the resulting accuracy–efficiency trade-off.

\section{Evaluation}
\label{sec:eval}

\subsection{Experimental Setup}
\label{sec:eval_setup}

\paragraph{Datasets}
We evaluate six public datasets.
CipherSpectrum~\cite{sok_enigma} contains 120,000 TLS~1.3 sessions from 40 domains.
We use 40,020 flows from 120 applications in CSTNET-TLS1.3~\cite{4394-fv34-24}.
The other four datasets provide legacy application and malicious-traffic benchmarks.
For ISCXTOR2016~\cite{icissp17} and ISCXVPN2016~\cite{icissp16vpn}, we use class-balanced subsets of 23,605 and 48,456 flows, respectively.
We select 10,000 flows from USTC-TFC2016~\cite{ustc-tfc2016} through class-stratified sampling.
This subset covers 20 benign and malware classes.
We also use 10,000 flows from CTU-13~\cite{10593100} for seven-class botnet traffic classification.

Datasets are split into an $8{:}1{:}1$ training-validation-test split.
CTU-13 uses capture-disjoint partitions with proportional sampling to preserve class proportions.
All samples from one capture remain in the same subset.
Input construction follows Sec.~\ref{sec:method_input}.

\paragraph{Baselines}
We compare NetCoFi with nine representative methods in Table~\ref{tab:performance_comparison}.
AppScanner~\cite{appscanner} identifies applications using statistical traffic features.
FlowPrint~\cite{flowprint} builds application fingerprints from temporal correlations among network flows.
These methods provide comparisons with feature-based and fingerprint-based classification.

The neural baselines cover several ways of modeling traffic.
Deep Packet~\cite{deeppacket} learns features directly from packet bytes.
FS-Net~\cite{fsnet} uses an encoder--decoder network to capture sequential patterns within flows.
TFE-GNN~\cite{tfegnn} constructs byte-level graphs and learns relationships between header and payload bytes.
ET-BERT~\cite{etbert} uses a bidirectional Transformer to learn contextual traffic representations.
YaTC~\cite{yatc} combines a multi-level flow representation with masked autoencoding.
NetMamba~\cite{netmamba} and PaT~\cite{pat} provide Mamba-based baselines for efficient sequence modeling.
The latter two help assess whether NetCoFi offers benefits beyond using an efficient backbone.

The main comparison trains all neural models from random initialization.
We also evaluate pretrained models in Table~\ref{tab:pretrain_compare}.
All models in this comparison use the same unlabeled pretraining corpus.
For NetCoFi, we pretrain the fine- and coarse-branch encoders.
The routing modules are randomly initialized before downstream multi-budget training.
This separates the comparison without pretraining from the evaluation of pretrained representations.

\paragraph{Metrics}
We report Accuracy (AC) and Macro-F1 (F1) for classification.
Efficiency is evaluated using inference latency, throughput, and peak GPU memory.
Throughput is reported in samples/s for model inference and Mb/s for the deployment pipeline.

\paragraph{Implementation Details}
Experiments use PyTorch 2.8.0 with CUDA 12.8 on NVIDIA GeForce RTX 4090 GPUs.
We train NetCoFi in FP32 with AdamW for 100 epochs.
The batch size is 64, the initial learning rate is $2.5\times10^{-4}$, and weight decay is 0.05.
The learning rate warms up linearly for the first 20 epochs.

Multi-budget training covers $\rho\in[0,1]$.
Each mini-batch uses both endpoints and two uniformly sampled intermediate budgets.
The full-budget pass at $\rho=1.0$ provides the distillation targets.
We set $T_{\mathrm{kd}}=2.0$, $\lambda_{\mathrm{kd}}=0.5$, and $\lambda_{\mathrm{budget}}=0.1$.
Routing budgets are specified in the corresponding experiments.

\subsection{Overall Performance}
\label{sec:eval_overall}
\begin{table*}[htbp]
\centering
\caption{Performance comparison of different methods across six datasets.}
\label{tab:performance_comparison}
\resizebox{\textwidth}{!}{%
\begin{tabular}{l|cc|cc|cc|cc|cc|cc}
\toprule
\multicolumn{1}{l|}{\multirow{2}{*}{Method}}
& \multicolumn{2}{c|}{ISCXTOR2016}
& \multicolumn{2}{c|}{ISCXVPN2016}
& \multicolumn{2}{c|}{USTC-TFC2016}
& \multicolumn{2}{c|}{CTU-13}
& \multicolumn{2}{c|}{CSTNET-TLS1.3}
& \multicolumn{2}{c}{CipherSpectrum} \\
\cmidrule{2-3} \cmidrule{4-5} \cmidrule{6-7} \cmidrule{8-9} \cmidrule{10-11} \cmidrule{12-13}
& AC & F1 & AC & F1 & AC & F1 & AC & F1 & AC & F1 & AC & F1 \\
\midrule
AppScanner~\cite{appscanner}      & 41.00 & 25.85 & 77.20 & 75.30 & 68.20 & 69.36 & 16.32 & 15.38 & 14.78 & 15.98 & 7.21  & 5.95 \\
FlowPrint~\cite{flowprint}        & 30.90 & 16.16 & 29.30 & 25.80 & 25.40 & 15.68 & 56.36 & 57.43 & 21.68 & 10.56 & 7.39  & 5.82 \\
\midrule
FS-Net~\cite{fsnet}               & 70.27 & 71.80 & 72.85 & 74.44 & 44.68 & 29.00 & 39.80 & 34.78 & 16.70 & 10.38 & 6.91  & 4.98 \\
TFE-GNN~\cite{tfegnn}             & 75.98 & 76.84 & 84.80 & 84.59 & 96.50 & 96.23 & 50.30 & 45.78 & 49.30 & 45.28 & 7.44  & 5.00 \\
DeepPacket~\cite{deeppacket}      & 72.56 & 71.32 & 85.98 & 83.31 & 93.24 & 93.78 & 53.87 & 43.95 & 30.21 & 30.16 & 8.32  & 8.11 \\
\midrule
ET-BERT~\cite{etbert}             & 92.11 & 92.76 & 90.76 & 84.35 & 94.30 & 94.23 & 70.70 & 69.65 & 54.07 & 52.87 & 30.27 & 30.98 \\
YaTC~\cite{yatc}                  & 90.24 & 90.27 & 87.91 & 86.93 & 95.69 & 95.41 & 80.65 & 69.56 & 60.33 & 55.33 & 45.28 & 40.33 \\
NetMamba~\cite{netmamba}          & 92.69 & 91.35 & 87.59 & 87.50 & 97.12 & 96.80 & 82.32 & 71.74 & 79.53 & 75.08 & 49.48 & 44.78 \\
PaT~\cite{pat}                    & 93.56 & 92.50 & 85.33 & 87.93 & \underline{\textbf{97.35}} & 96.39 & 83.16 & 73.11 & 81.22 & 80.78 & 48.97 & 44.29 \\
\midrule
NetCoFi                           & \underline{\textbf{94.13}} & \underline{\textbf{92.81}} & \underline{\textbf{91.95}} & \underline{\textbf{91.92}} &  97.14 & \underline{\textbf{97.01}} & \underline{\textbf{85.20}} & \underline{\textbf{74.79}} & \underline{\textbf{85.37}} & \underline{\textbf{82.75}} & \underline{\textbf{51.19}} & \underline{\textbf{45.16}} \\
\bottomrule
\end{tabular}}
\end{table*}

Table~\ref{tab:performance_comparison} compares classification performance without pretraining.
NetCoFi achieves the highest Macro-F1 on all six datasets, also the highest Accuracy on five datasets.

On CSTNET-TLS1.3, NetCoFi outperforms PaT by 4.15 percentage points in Accuracy and 1.97 points in Macro-F1.
On CTU-13, the corresponding gains over PaT are 2.04 and 1.68 points.
The improvements cover both TLS application identification and botnet traffic classification.
They also appear in both overall accuracy and class-averaged performance.
NetCoFi therefore combines budget flexibility with strong classification performance.

\begin{table*}[htbp]
\centering
\caption{Pretrained performance comparison across six datasets.}
\label{tab:pretrain_compare}
\setlength{\tabcolsep}{4pt}
\renewcommand{\arraystretch}{1.0}
\resizebox{\textwidth}{!}{
\begin{tabular}{l|cc|cc|cc|cc|cc|cc}
\toprule
\multicolumn{1}{l|}{\multirow{2}{*}{Method}}
& \multicolumn{2}{c|}{ISCXTOR2016}
& \multicolumn{2}{c|}{ISCXVPN2016}
& \multicolumn{2}{c|}{USTC-TFC2016}
& \multicolumn{2}{c|}{CTU-13}
& \multicolumn{2}{c|}{CSTNET-TLS1.3}
& \multicolumn{2}{c}{CipherSpectrum} \\
\cmidrule(lr){2-3}
\cmidrule(lr){4-5}
\cmidrule(lr){6-7}
\cmidrule(lr){8-9}
\cmidrule(lr){10-11}
\cmidrule(lr){12-13}
& AC & F1 & AC & F1 & AC & F1 & AC & F1 & AC & F1 & AC & F1 \\
\midrule
ET-BERT~\cite{etbert}
& 92.78 & 92.85
& 92.95 & 91.98
& \underline{\textbf{98.10}} & \underline{\textbf{98.22}}
& 70.90 & 70.55
& 60.54 & 59.07
& 50.99 & 44.99 \\

YaTC~\cite{yatc}
& 95.22 & 94.44
& 93.16 & 93.18
& 97.67 & 97.52
& 82.92 & 72.86
& 88.98 & 87.61
& 62.91 & 50.99 \\

NetMamba~\cite{netmamba}
& 95.01 & 94.11
& 94.00 & 93.89
& 97.86 & 97.94
& 85.30 & 74.31
& 89.98 & 89.78
& 70.89 & 60.78 \\

PaT~\cite{pat}
& 95.12 & 94.31
& \underline{\textbf{94.50}} & 94.29
& 97.31 & 97.27
& 85.31 & 73.98
& \underline{\textbf{91.23}} & \underline{\textbf{90.50}}
& 71.00 & 61.20 \\
\midrule

NetCoFi
& \underline{\textbf{95.44}} & \underline{\textbf{94.48}}
& 94.38 & \underline{\textbf{94.40}}
& 97.54 & 97.38
& \underline{\textbf{85.60}} & \underline{\textbf{74.51}}
& 90.96 & 89.35
& \underline{\textbf{71.90}} & \underline{\textbf{61.69}} \\
\bottomrule
\end{tabular}
}
\end{table*}

Table~\ref{tab:pretrain_compare} reports results after pretraining on the same unlabeled corpus.
NetCoFi achieves the highest Macro-F1 on four of the six datasets.
It also leads in Accuracy on ISCXTOR2016, CTU-13, and CipherSpectrum.
On CipherSpectrum, pretraining raises its Macro-F1 from 45.16\% to 61.69\%.
This shows that NetCoFi can benefit from pretrained representations.

ET-BERT leads on USTC-TFC2016, while PaT leads on CSTNET-TLS1.3.
NetCoFi remains competitive in this setting, although its advantage is more consistent without pretraining.

\subsection{Budget-Conditioned Inference and Efficiency}
\label{sec:budget_conditioned_efficiency}
We examine how the routing budget affects classification performance and inference cost. We also compare training strategies and measure token counts, throughput, and GPU memory usage. We first show the resulting accuracy--latency frontier, then check budget fidelity, analyze the effect of sandwich-style multi-budget training, and finally report the speed--memory operating points.

\begin{figure*}[htbp]
    \centering
    \includegraphics[width=\textwidth]{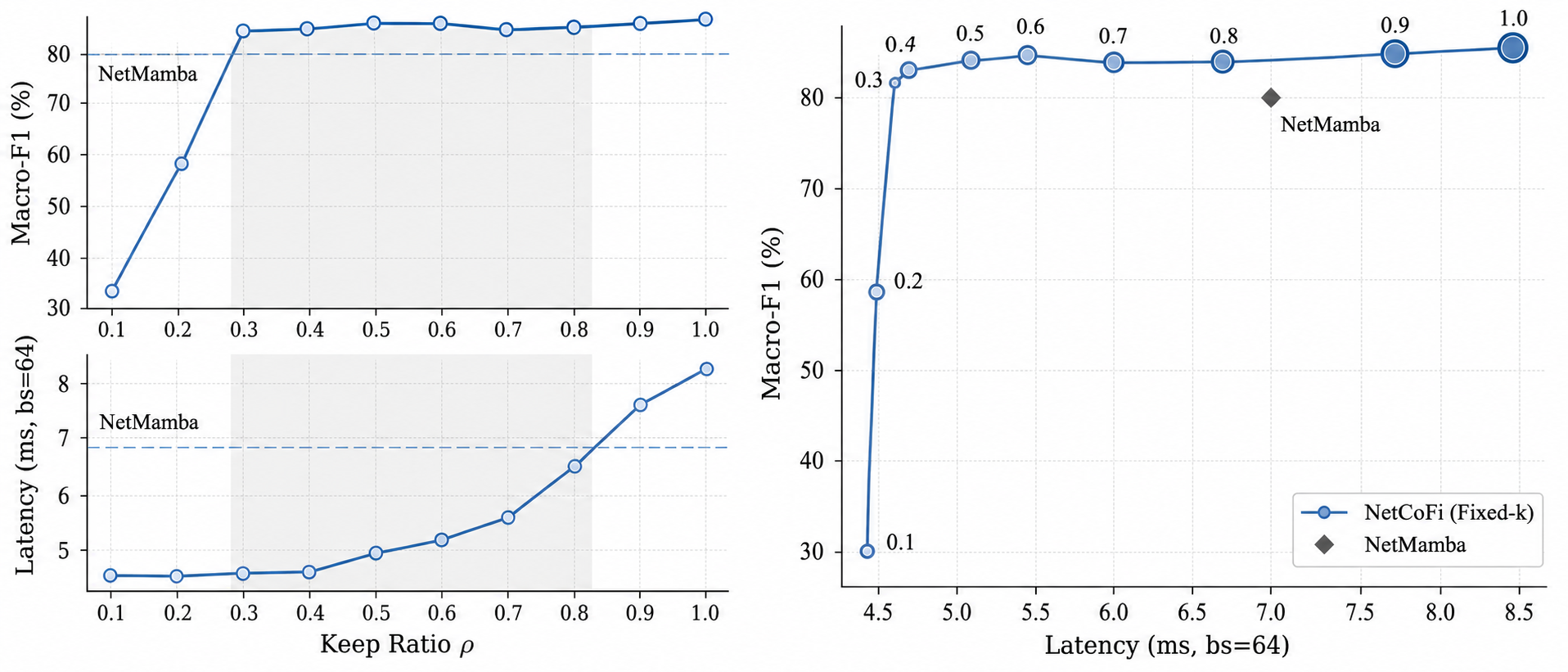}
    \caption{Budget behavior and operating frontier of NetCoFi. \textbf{(a,~b)} Budget-response curves under deterministic fixed-$k$ seed routing: Macro-F1~(a) and latency~(b) versus routing budget~$\rho$, with NetMamba as a horizontal reference in both panels. \textbf{(c)} Operating points in the latency--Macro-F1 plane}
    \label{fig:budget_frontier}
\end{figure*}

\subsubsection{Accuracy--Latency Trade-off}
\label{sec:accuracy_latency_frontier}
Figure~\ref{fig:budget_frontier} shows Macro-F1 and batch latency at different routing budgets.
NetMamba provides a fixed reference, while NetCoFi offers several choices with the same trained model.
Macro-F1 rises sharply as $\rho$ increases from 0.1 to 0.3.
Further budget increases bring smaller changes in Macro-F1 but generally increase latency.

At the tested budgets between $\rho=0.3$ and $0.8$, NetCoFi achieves higher Macro-F1 and lower batch latency than NetMamba.
Reducing the budget further causes a substantial loss in classification performance.
These results identify a useful middle range where NetCoFi reduces inference cost while retaining strong classification performance.

\subsubsection{Budget Calibration and Budget Fidelity}
\label{sec:budget_calibration}
We next inspect the token sequences passed to the two encoders.
The fine-token count includes the gathered neighborhoods after expansion, rather than only unique token positions.
The coarse-token count is $C_{\mathrm{eff}}$, and the total is the sum of both branch lengths.

\begin{figure}[htbp]
    \centering
    \includegraphics[width=0.84\linewidth]{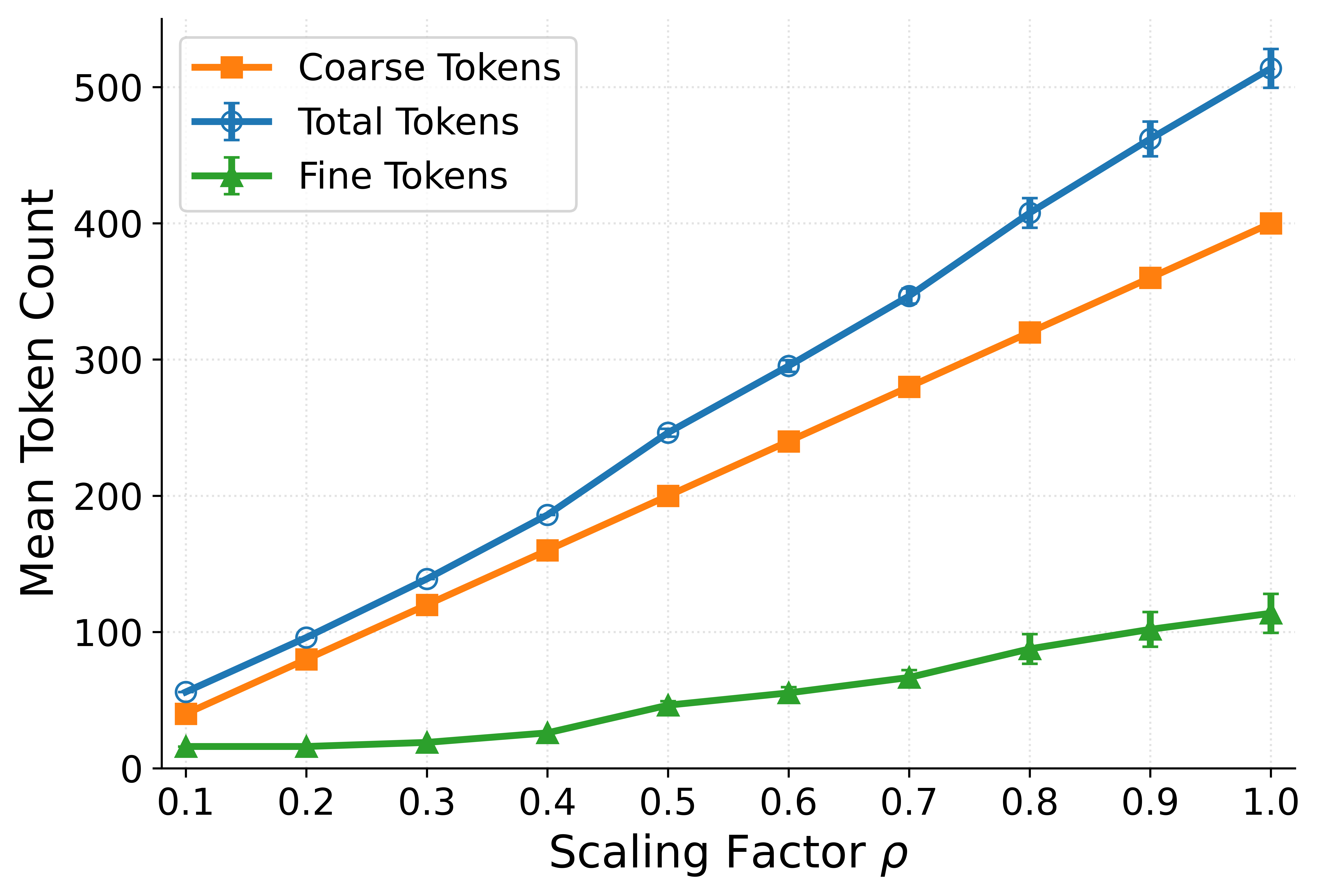}
    \caption{Budget calibration curves under the current fixed-$k$ implementation. As the routing budget $\rho$ increases, the mean coarse-token count grows nearly linearly, while the mean fine-token count and mean total token count increase smoothly and monotonically.}
    \label{fig:budget_calib}
\end{figure}

Figure~\ref{fig:budget_calib} shows that the coarse-token count grows nearly linearly with $\rho$.
The fine and total counts also increase as the budget rises.
The fine-token count reflects both seed selection and neighborhood expansion.

\begin{figure}[htbp]
    \centering
    \includegraphics[width=\linewidth]{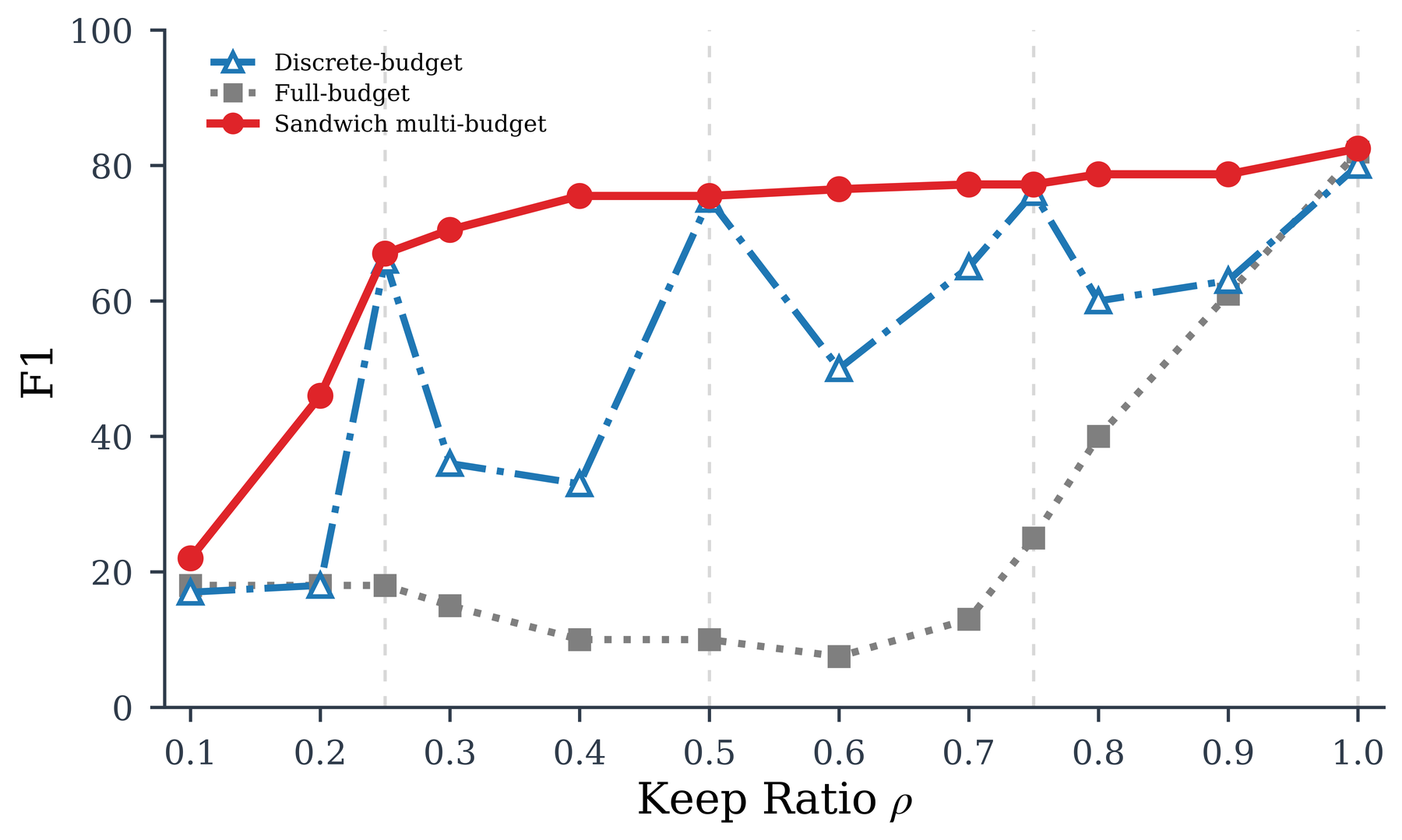}
    \caption{Effect of the budget-training strategy on performance across the evaluated routing budgets.}
    \label{fig:multibudget_effect}
\end{figure}

\subsubsection{Effect of Multi-Budget Training}
\label{sec:multibudget_training_effect}
Figure~\ref{fig:multibudget_effect} compares three training strategies.
Full-budget training uses only $\rho=1.0$.
Discrete-budget training uses $\{0.25,0.5,0.75,1.0\}$.
Sandwich-style training uses both endpoints and random intermediate budgets in each update.

Full-budget training loses substantial Macro-F1 when the Routing budget is reduced.
Discrete-budget training performs better near its training budgets but shows sharp drops at intermediate values.
Sandwich-style training produces a smoother curve and stronger performance at reduced budgets.
This shows why training across different budgets is useful: the model must learn to classify with different amounts of retained detail.

\subsubsection{Speed--Memory Trade-off}
\label{sec:speed_memory_tradeoff}
We measure latency and throughput at batch size 64 after warm-up.
Repeated forward passes use CUDA synchronization.
The measurements include token selection and routing but exclude data loading.

\begin{figure}[htbp]
    \centering
    \includegraphics[width=0.82\linewidth]{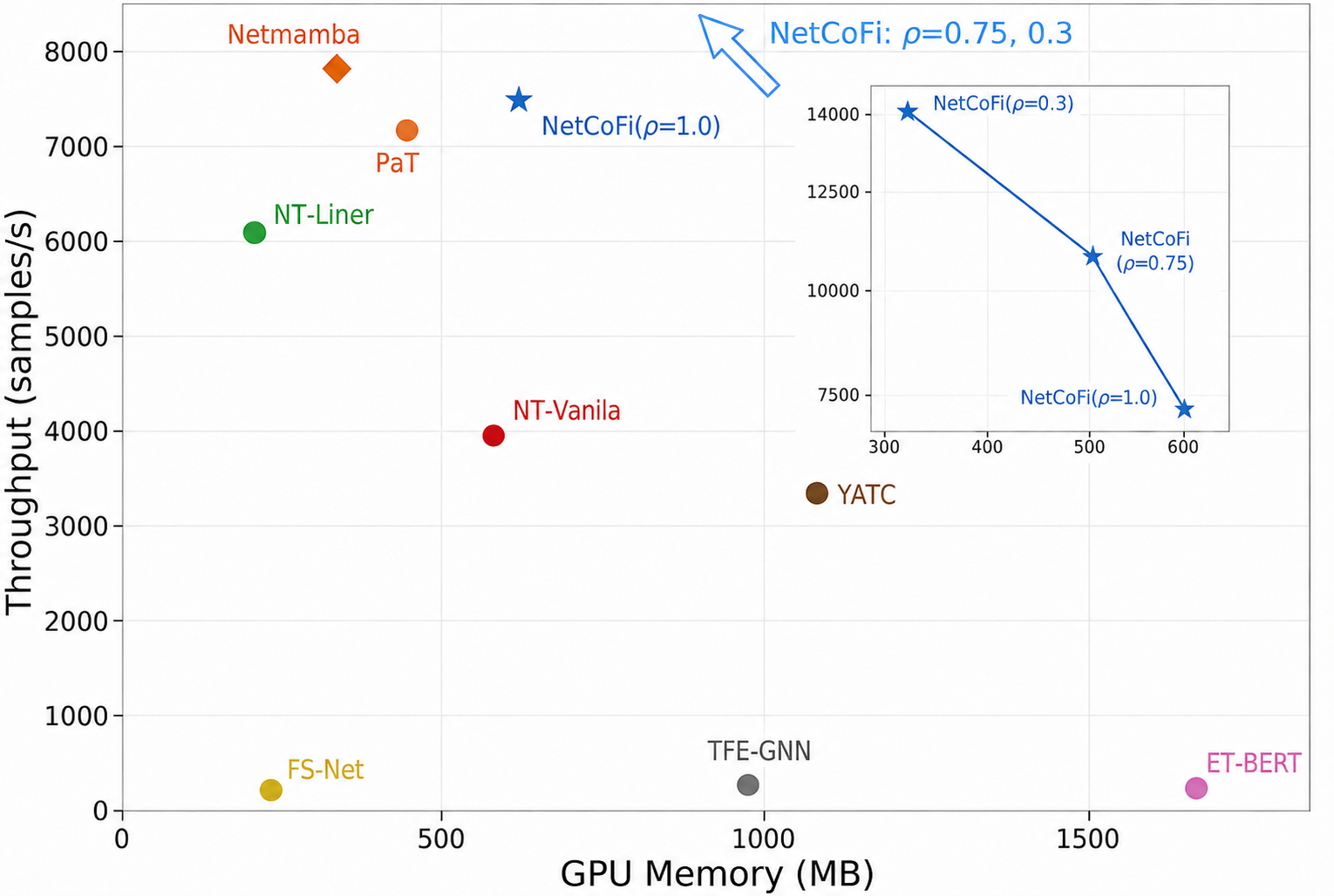}
    \caption{Inference speed and GPU memory usage at batch size $64$. NetCoFi is shown at multiple routing budgets, while competing methods appear as static operating points.}
    \label{fig:speed_memory}
\end{figure}

At $\rho=1.0$, NetCoFi has throughput close to NetMamba but uses more GPU memory.
Reducing $\rho$ increases throughput and lowers memory usage, as shown in Fig.~\ref{fig:speed_memory}.
The settings at $\rho=0.75$ and $0.3$ illustrate this change with the same trained model.

\subsection{Adaptive vs. Static Token Usage}
\label{sec:same_depth_token_reduction}
We compare NetCoFi with a static token-reduction version of NetMamba.
Both use the same backbone depth.
The static baseline uses a fixed 120-byte input, while NetCoFi uses $\rho=0.3$.
This comparison tests adaptive routing with coarse context against fixed input reduction.

\begin{table}[htbp]
\centering
\caption{Comparison with a static token-reduction baseline in the low-budget regime.}
\label{tab:same_depth_token_reduction}
\setlength{\tabcolsep}{4pt}
\renewcommand{\arraystretch}{0.95}
\footnotesize
\begin{tabular}{lcc}
\toprule
Dataset & NetMamba & NetCoFi \\
\midrule
ISCXVPN2016   & 82.30 & \textbf{86.19} \\
CTU-13        & 67.43 & \textbf{72.69} \\
CSTNET-TLS1.3 & 69.15 & \textbf{83.57} \\
\midrule
Average       & 72.96 & \textbf{80.82} \\
\bottomrule
\end{tabular}
\par\vspace{2pt}
\begin{minipage}{0.88\linewidth}
\scriptsize
\emph{Note} NetMamba uses a static 120B token-reduction baseline; NetCoFi is evaluated at $\rho=0.3$.
\end{minipage}
\end{table}

Table~\ref{tab:same_depth_token_reduction} shows that NetCoFi outperforms the static baseline on all three datasets.
Macro-F1 improves by 3.89 percentage points on ISCXVPN2016, 5.26 on CTU-13, and 14.42 on CSTNET-TLS1.3.
The average rises from 72.96\% to 80.82\%, a gain of 7.86 percentage points.
These results favor NetCoFi's selective fine-grained processing with coarse context over a fixed shortened input.

\subsection{Structural Ablation}
\label{sec:ablation}
We perform controlled structural ablations to identify how the main components contribute to NetCoFi. All variants in Table~\ref{tab:ablation_grouped_ac_f1} use the same data split and training protocol. To avoid conflating structural effects with compute allocation, all ablations use the same budget setting ($\rho=1.0$), and each variant changes only the target component while keeping the other hyperparameters fixed.

\begin{table}[t]
\centering
\caption{Structural ablation of NetCoFi on two representative datasets.}
\label{tab:ablation_grouped_ac_f1}

\renewcommand{\arraystretch}{1.10}
\setlength{\tabcolsep}{4pt}
\footnotesize

\begin{tabular*}{\columnwidth}{@{\extracolsep{\fill}} l cc|cc @{}}
\toprule
\multirow{2}{*}{Variant} &
\multicolumn{2}{c|}{CSTNET-TLS1.3} &
\multicolumn{2}{c}{CTU-13} \\
\cmidrule(lr){2-3}
\cmidrule(lr){4-5}
& AC & F1 & AC & F1 \\
\midrule
Full NetCoFi  & \textbf{85.37} & \textbf{82.75}
               & \textbf{85.20} & \textbf{74.79} \\
Fine-only     & 83.00 & 80.29 & 82.10 & 69.18 \\
Coarse-only   & 74.28 & 70.73 & 76.07 & 68.02 \\

\midrule
Fixed ROI      & 78.02 & 74.97 & 76.27 & 68.03 \\
Random ROI     & 75.99 & 73.19 & 75.67 & 67.71 \\
No SideFeature & 79.38 & 74.37 & 74.38 & 66.59 \\
Uniform WPool  & 74.28 & 71.69 & 76.17 & 67.40 \\
No Expand      & 80.75 & 77.85 & 82.62 & 72.98 \\

\midrule
No CrossGate   & 83.75 & 81.19 & 84.41 & 74.45 \\

\bottomrule
\end{tabular*}

\vspace{2pt}

\end{table}

\subsubsection{Core Coarse-Fine Dual-Branch Architecture}
We first examine the contribution of the coarse-fine dual-branch architecture.
As shown in Table~\ref{tab:ablation_grouped_ac_f1}, the full model outperforms both single-branch variants on CSTNET-TLS1.3 and CTU-13.
Fine-only consistently performs better than Coarse-only, showing that fine-grained token modeling remains the main discriminative pathway.
Adding the coarse branch further improves Macro-F1 by 2.46 percentage points on CSTNET-TLS1.3 and 5.61 points on CTU-13.
This indicates that compressed global context provides complementary information to the fine branch.

\subsubsection{Routing and Context Construction}
The second group studies whether the routing pipeline matters beyond sparse token selection. Fixed ROI and Random ROI degrade performance relative to the full model, supporting the need for data-adaptive routing in contrast to a fixed or random selection rule. No Expand remains a relatively strong variant but still trails the full model, suggesting that top-ranked seed tokens benefit from nearby context. Uniform WPool also degrades performance, indicating that the coarse branch is more useful when it preserves residual context through anti-importance weighted pooling rather than uniform compression. These results suggest that informative scoring, deterministic fixed-$k$ seed routing, local neighborhood construction, and structured coarse retention jointly contribute to the final model.

\begin{table}[htbp]
\centering
\caption{Per-flow split results under packet-level and flow-level classification tasks.}
\label{tab:flow_based_classification}
\setlength{\tabcolsep}{2pt}
\renewcommand{\arraystretch}{0.92}
\scriptsize
\resizebox{\linewidth}{!}{%
\begin{tabular}{l|cc|cc|cc|cc}
\toprule
\multirow{3}{*}{\textbf{\begin{tabular}[c]{@{}c@{}}Model\\ (Per-flow split)\end{tabular}}}
& \multicolumn{4}{c|}{\textbf{VPN-app (16)}}
& \multicolumn{4}{c}{\textbf{TLS-120}} \\
\cmidrule(lr){2-5} \cmidrule(lr){6-9}
& \multicolumn{2}{c|}{\textbf{Packet-level}}
& \multicolumn{2}{c|}{\textbf{Flow-level}}
& \multicolumn{2}{c|}{\textbf{Packet-level}}
& \multicolumn{2}{c}{\textbf{Flow-level}} \\
\cmidrule(lr){2-3} \cmidrule(lr){4-5} \cmidrule(lr){6-7} \cmidrule(lr){8-9}
& \textbf{AC} & \textbf{F1}
& \textbf{AC} & \textbf{F1}
& \textbf{AC} & \textbf{F1}
& \textbf{AC} & \textbf{F1} \\
\midrule
ET-BERT       & 82.1 & 68.6 & 57.4 & 52.5 & 27.1 & 20.8 & 54.2 & 50.5 \\
YaTC          & 78.9 & 63.2 & 58.3 & 52.6 & 33.0 & 26.2 & 77.3 & 74.8 \\
NetMamba      & 79.6 & 65.2 & 52.0 & 48.1 & 33.8 & 29.1 & 73.8 & 71.1 \\
\midrule
\textit{NetCoFi} & 81.6 & 67.6 & 53.5 & 49.3 & 37.5 & 32.4 & 78.2 & 75.7 \\
\bottomrule
\end{tabular}
}
\end{table}

\subsubsection{CrossGate}
The third group evaluates whether explicit interaction between the two branches is useful. Removing CrossGate causes moderate performance drops on both datasets. Macro-F1 decreases by 1.56 percentage points on CSTNET-TLS1.3 and 0.34 points on CTU-13. This suggests that the fine branch benefits from modulation by coarse contextual summaries, while also indicating that CrossGate is one contributing component.

\begin{figure*}[htbp]
    \centering
    \includegraphics[width=0.90\textwidth]{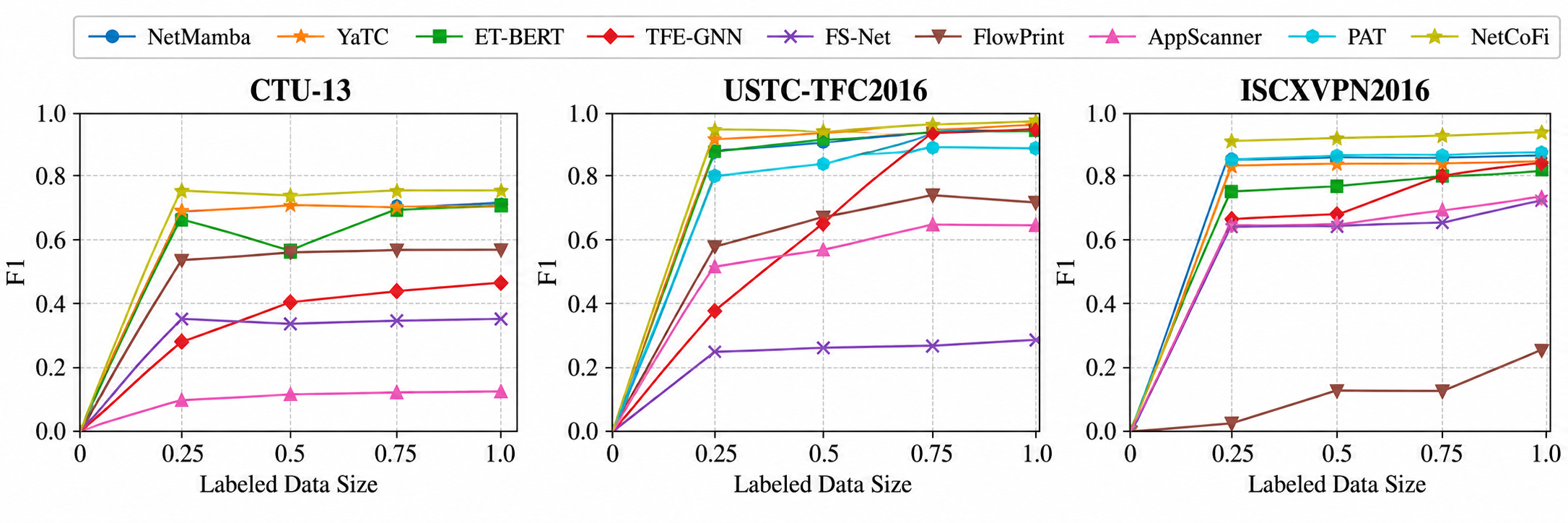}
    \caption{Few-shot performance curves under different labeled data ratios.}
    \label{fig:fewshot}
\end{figure*}

\begin{figure}[htbp]
\centering
\includegraphics[width=0.72\linewidth]{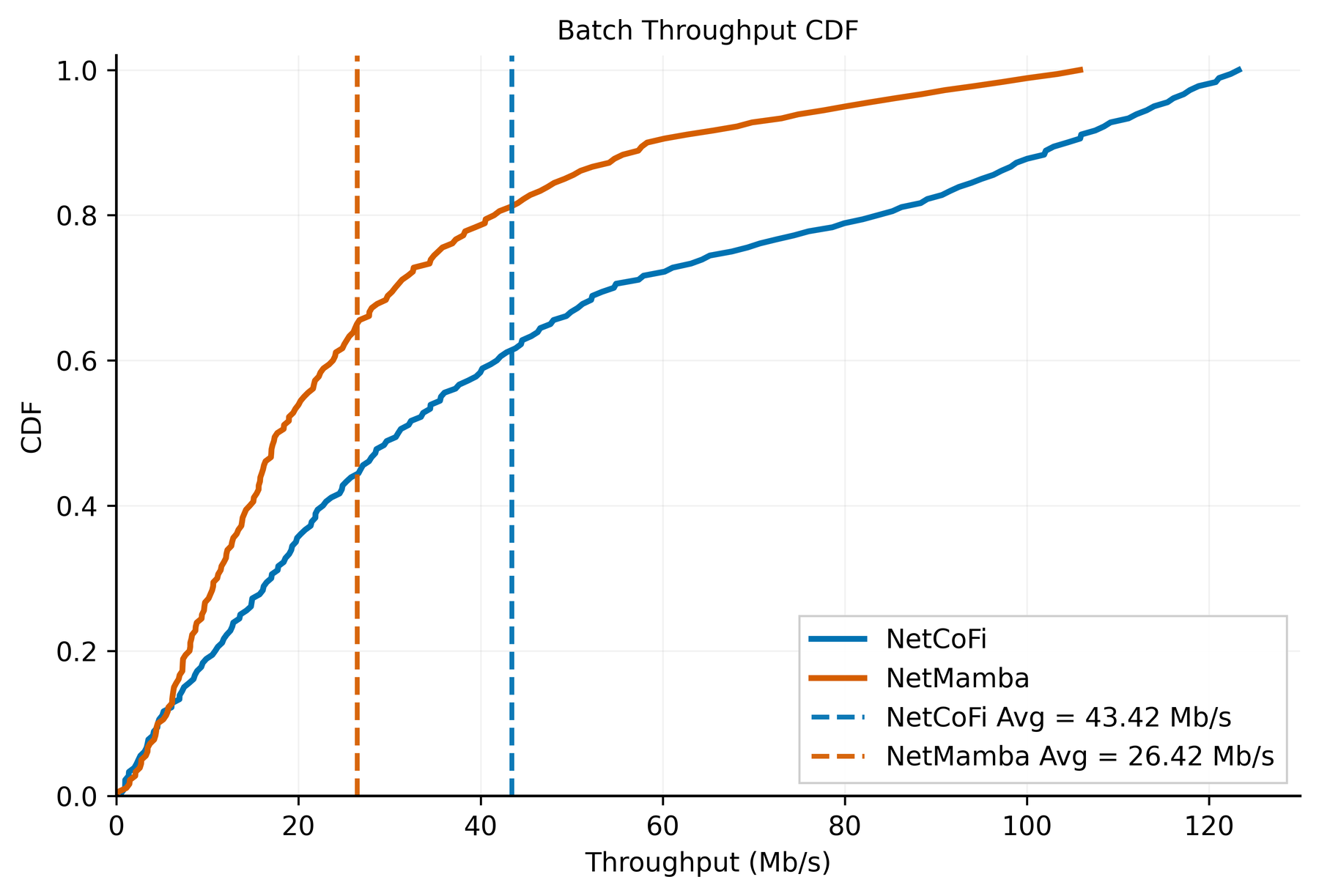}
\caption{Empirical CDF of batch throughput in the DPDK--GPU pipeline.
Dashed lines indicate mean batch throughput.}
\label{fig:deployment_throughput_cdf}
\end{figure}

\subsection{Analysis under Strict Split Strategies}
\label{sec:strict_split}
In encrypted traffic classification, the split strategy has a direct effect on how reliable the evaluation is. Packet-level splitting can assign packets from the same session to different subsets, allowing a model to exploit fields shared within a session or highly correlated across packets from the same session. This can introduce shortcut learning. Per-flow split is stricter because all packets from a given flow are assigned exclusively to one subset, making the evaluation closer to real-world scenarios on unseen flows.

VPN-app (16) and TLS-120 are auxiliary benchmarks used only in this strict-split analysis, not additional datasets in the main six-benchmark comparison. They correspond to a 16-class VPN application task and a 120-class TLS application task, respectively. To evaluate generalization under this stricter protocol, we construct both flow-level and packet-level classification tasks on VPN-app and TLS-120, and adopt flow-level splitting as the primary isolation strategy. The flow is the unit of data isolation. Both a flow and its constituent packets inherit the same label. In packet-level classification, each training sample is still a single packet, but the training, validation, and test sets are first partitioned at the flow level so that packets from the same flow appear in only one subset. This prevents explicit or implicit flow identifiers from leaking between training and test data. For flow-based models such as NetMamba, in the packet-level task we duplicate each packet to form a pseudo-flow of length five, matching the original input format. Following NetMamba's preprocessing, we also clear IP address and port fields to weaken explicit flow identifiers. For the flow-level task, we retain only flows with at least five packets and use the first five packets as input, consistent with the original flow encoder.

During splitting, we first construct training and test sets at the flow level while balancing the distribution of long flows across subsets as much as possible. The test set preserves the class distribution induced by the flow-level split, whereas the training and validation sets are balanced through under-sampling to reduce the influence of majority classes during training. We also construct the token-level and global auxiliary features required by NetCoFi for packet samples.

Table~\ref{tab:flow_based_classification} shows that NetCoFi outperforms NetMamba on both TLS-120 tasks under per-flow splits. The Macro-F1 gains are 3.3 points for packet-level classification and 4.6 points for flow-level classification. These results show that the gains persist when test flows are unseen during training.

\subsection{Few-shot Evaluation}
\label{sec:fewshot_eval}
Figure~\ref{fig:fewshot} examines label efficiency by plotting F1 against the fraction of labeled training data on CTU-13, USTC-TFC2016, and ISCXVPN2016. The main question is whether the proposed coarse-fine dual-branch routing remains effective when supervision is scarce, unlike only in the fully supervised regime.

The curves show that NetCoFi remains competitive or strongest across labeled-data ratios, especially on CTU-13. NetCoFi remains near the top across all three datasets and all labeled-data ratios, with particularly consistent gains in low-data settings.

\subsection{DPDK--GPU Deployment}
\label{sec:real_world_deployment}
We test NetCoFi in a traffic processing pipeline with DPDK packet capture and a Python/GPU backend.
A flow assembler groups packets using a canonical bidirectional five-tuple.
Packets from both directions share one flow state, while direction is retained as a side feature.
For each flow, the assembler records packet bytes, length, direction, TCP flags, and inter-arrival time.
Once five packets are available, it constructs the same fixed-length input used in the main experiments.
Flow windows are then batched for GPU inference without waiting for the session to end.

NetCoFi and NetMamba share the same DPDK front-end, flow assembler, feature processing, batching policy, and hardware.
We measure batch throughput from DPDK-delivered packets to model predictions.
This includes flow aggregation, feature construction, routing where applicable, and GPU inference.
Disk I/O and offline trace loading are excluded.

Figure~\ref{fig:deployment_throughput_cdf} shows the batch throughput distribution, and it demonstrates that NetCoFi improves throughput consistently across the batch distribution, as evidenced by the rightward shift of the CDF curve.
Mean batch throughput rises from 26.42~Mb/s to 43.42~Mb/s, making it 1.64 times that of NetMamba.

\section{Conclusion}
Encrypted traffic classification requires efficient inference under limited inference resources, together with reliable discrimination under limited payload visibility. We presented \methodname, a coarse-fine dual-branch Mamba architecture that learns token importance to route informative stride tokens to a fine branch while pooling complementary context into a coarse branch. By introducing an explicit routing budget, which is resolved into stable seed, neighborhood, and coarse-token routing variables, and adopting a sandwich-style multi-budget training strategy with endpoint and random-intermediate budget sampling, \methodname{} supports more stable budget-conditioned inference within the trained budget interval and outperforms NetMamba across multiple public benchmarks.

\bibliographystyle{IEEEtran}
\bibliography{references}

\end{document}